\documentclass[nofootinbib,amsmath,amssymb,a4paper]{revtex4}

\usepackage{graphicx}
\usepackage{bm}

\begin{document}

\title{Quenched lattice calculation of semileptonic heavy-light meson form factors}

\author{G.M. de Divitiis$^{a,b}$, R. Petronzio$^{a,b}$, N. Tantalo$^{b,c}$}
\affiliation{\vskip 10pt
$^{a}$~Universit\`a di Roma ``Tor Vergata'', I-00133 Rome, Italy\\
$^{b}$~INFN sezione di Roma ``Tor Vergata'', I-00133 Rome, Italy\\
$^{c}$~Centro Enrico Fermi, I-00184 Rome, Italy
}%

\begin{abstract}
We calculate, in the continuum limit of quenched lattice QCD, the matrix elements of the
heavy-heavy vector current between heavy-light pseudoscalar meson states. 
We present the form factors for different values of the initial and final meson masses 
at finite momentum transfer. 
In particular, we calculate the non-perturbative correction to the differential decay rate of
the process $B\rightarrow D\ell\nu$ including the case of a non-vanishing lepton mass. 
\end{abstract}

\maketitle

\section{Introduction}
Semileptonic decays of heavy-light mesons play a central role in the study of flavour physics
both on the experimental and theoretical sides.
The extraction of the Cabibbo--Kobayashi--Maskawa~\cite{Cabibbo:1963yz,Kobayashi:1973fv} 
matrix element $V_{cb}$, for example, requires the experimental measurement of the decay rate 
of the process $B\rightarrow D^{(*)}\ell\nu_{\ell}$ and the theoretical calculation of the hadron
matrix elements of the flavour changing weak currents. A non-perturbative estimate of
the matrix elements can be obtained by lattice QCD.
Furthermore, within the heavy quark effective theory (HQET) it has been 
shown~\cite{Isgur:1989ed} that the semileptonic transitions between heavy-light mesons
can be parametrized, at leading order of 
the expansion in the inverse heavy quark mass, 
in terms of a universal form factor known as Isgur-Wise function.
The Isgur-Wise function is universal in the sense that it describes any semileptonic
decay mediated by heavy-heavy weak currents regardless of the flavour of the initial
and final heavy quarks and of the spins of the mesons. From the phenomenological point
of view it is relevant to know the size of the corrections to the Isgur-Wise limit and
to establish at which order the heavy quark expansion has to be truncated to
produce useful results down to the charm mass.

Matrix elements of the vector heavy-heavy currents between pseudoscalar
meson states are parametrized in terms of two form factors.
In the case of the light leptons $\ell=e,\mu$,
the differential decay rate of the process $B\rightarrow D\ell\nu_{\ell}$ is proportional
to the square of a particular linear combination of the two, $G^{B\rightarrow D}$. 
The BaBar and Belle collaborations have already measured~\cite{Aubert:2007ab,Matyja:2007kt} 
the branching ratios of the processes $B\rightarrow D^{(*)}\tau\nu_{\tau}$ and
a future measurement of the differential decay rate will make possible to extract
$V_{cb}$ also from this channel.
In this case a separate knowledge of the form factors
is required, both in the Standard Model and in its minimal extensions 
(see for example refs.~\cite{Kiers:1997zt,Chen:2006nua}).
In ref.~\cite{de Divitiis:2007ui} we have already shown our final results for $G^{B\rightarrow D}$
by focusing on their phenomenological implications without giving all the details of the calculation.
Here we present detailed results of continuum and chiral extrapolations and separate
estimates of the two independent form factors for several values of the initial and final
heavy quark masses together with an analysis of the infinite heavy quark mass limit.
In addition, we make a prediction for the ratio of the differential decay rates
of the processes $B\rightarrow D\ell\nu_{\ell}$ with $\ell=\tau$ and $\ell=e,\mu$.   

The simulation of relativistic heavy quarks with masses ranging from the physical 
$b$ mass down to the physical $c$ mass has been performed by using  
the step scaling method (SSM)~\cite{Guagnelli:2002jd}, already applied 
successfully to the determination of heavy 
quark masses and heavy-light meson
decay constants~\cite{deDivitiis:2003wy,deDivitiis:2003iy,Guazzini:2006bn}. 
The SSM allows to reconcile large quark masses with 
adequate lattice resolution and large physical volumes.
The two form factors have been calculated for different values of the momentum transfer by making 
use of flavour twisted boundary conditions~\cite{deDivitiis:2004kq}, 
that shift the discretized set of lattice momenta by an arbitrary amount 
(see also~\cite{Bedaque:2004kc,Sachrajda:2004mi,Flynn:2005in}).

The plan of the paper is as follows. In section~\ref{sec:ff} we introduce the form factors
in the continuum theory and re-derive the Luke's theorem~\cite{Luke:1990eg}.  
In sections~\ref{sec:notations} and~\ref{sec:ssm} we set up the lattice notation 
and describe the calculation. 
In sections~\ref{sec:fvr} we discuss the results at finite volumes
while in section~\ref{sec:fr} we show our final results. We draw our conclusions
in section~\ref{sec:conc}.

\section{Form factors} \label{sec:ff}
The semileptonic decay of a pseudoscalar meson into another pseudoscalar meson is mediated by
the vector part of the weak $V-A$ current. The corresponding matrix element can be 
parametrized in terms of two form factors,
\begin{eqnarray}
\langle \mathcal{M}_f\vert \ V^\mu \ \vert \mathcal{M}_i\rangle=
(p_i+p_f)^\mu \ f_+^{i\rightarrow f} + (p_i-p_f)^\mu \ f_-^{i\rightarrow f}
\nonumber
\end{eqnarray}
or, equivalently, 
\begin{eqnarray}
\frac{\langle \mathcal{M}_f\vert \ V^\mu \ \vert \mathcal{M}_i\rangle}{\sqrt{M_i M_f}}=
(v_i+v_f)^\mu \ h_+^{i\rightarrow f} + (v_i-v_f)^\mu \ h_-^{i\rightarrow f}
\label{eq:hphmdef}
\end{eqnarray}
where $v_{i,f}=p_{i,f}/{M_{i,f}}$ are the $4$-velocities of the mesons. 
The relations between the $h_{\pm}^{i\rightarrow f}$ and the $f_\pm^{i\rightarrow f}$
parametrizations are given by
\begin{eqnarray}
h_\pm^{i\rightarrow f}=\frac{(M_i+M_f)f_\pm^{i\rightarrow f}+
(M_i-M_f)f_\mp^{i\rightarrow f}}{2\sqrt{M_i M_f}}
\nonumber \\ \nonumber \\ \nonumber \\
f_\pm^{i\rightarrow f}=\frac{(M_i+M_f)h_\pm^{i\rightarrow f}
-(M_i-M_f)h_\mp^{i\rightarrow f}}{2\sqrt{M_i M_f}}
\nonumber
\end{eqnarray}
In the rest of this paper we work in the $h_\pm^{i\rightarrow f}$ parametrization that,
as it will emerge from the discussion below, is more convenient for the 
study of the dependence of the form factors upon the masses of the initial and final heavy 
quarks.

The form factors depend upon the masses of the parent and daughter particles and upon
$w\equiv v_f\cdot v_i$
\begin{eqnarray}
h_\pm^{i\rightarrow f}(w)\equiv h_\pm(w,M_i,M_f),
\nonumber
\end{eqnarray}
Time reversal and hermiticity imply that $h_+^{i\rightarrow f}$ and
$h_-^{i\rightarrow f}$ are real. 
Furthermore they imply that $h_+^{i\rightarrow f}$ is even under the 
interchange of the initial and final states while $h_-^{i\rightarrow f}$ is odd,
\begin{eqnarray}
h_+(w,M_i,M_f)\, =\,h_+(w,M_f,M_i), \qquad \qquad 
h_-(w,M_i,M_f)\, =\,-\ h_-(w,M_f,M_i)
\label{eq:evenodd}
\end{eqnarray}
In eq.~(\ref{eq:hphmdef}) one can consider the limit in which both 
meson masses go to infinity at fixed $4$-velocity; 
the left hand side is well defined in this limit and, consequently, 
also the form factors.
It is thus legitimate to make a change of variables from the meson masses to the parameters 
$\varepsilon_+$ and $\varepsilon_-$, defined as
\begin{eqnarray}
\varepsilon_+=\frac{1}{M_f}+\frac{1}{M_i},
\qquad \qquad
\varepsilon_-=\frac{1}{M_f}-\frac{1}{M_i}
\nonumber
\end{eqnarray}
Expressed as functions of the new variables, 
$h^{i\rightarrow f}_\pm\equiv h_\pm(w,\varepsilon_+,\varepsilon_-)$ 
are well defined at $\varepsilon_+=0$ and $\varepsilon_-=0$ and can be
expanded in power series around these points. 
The symmetry properties of eq.~(\ref{eq:evenodd}) force the odd(even) powers of 
$\varepsilon_-$ to vanish into the expansion of 
$h_+^{i\rightarrow f}$ ($h_-^{i\rightarrow f}$), i.e.
\begin{eqnarray}
h_+(w,\varepsilon_+,\varepsilon_-) &=& h_+(w,0,0) + 
\varepsilon_+ \frac{\partial h_+(w,0,0)}{\partial \varepsilon_+} +
\frac{\varepsilon_+^2}{2} \frac{\partial^2 h_+(w,0,0)}{\partial \varepsilon_+^2} +
\frac{\varepsilon_-^2}{2} \frac{\partial^2 h_+(w,0,0)}{\partial \varepsilon_-^2} + \dots
\nonumber \\ \nonumber \\
h_-(w,\varepsilon_+,\varepsilon_-) &=& 
\varepsilon_- \frac{\partial h_-(w,0,0)}{\partial \varepsilon_-} +
\frac{\varepsilon_- \varepsilon_+}{2} \frac{\partial^2 h_-(w,0,0)}
{\partial \varepsilon_- \partial \varepsilon_+} + \dots
\label{eq:taylorexp1}
\end{eqnarray}
In the elastic case, when the initial and final mesons coincide,
$h_-^{i\rightarrow i}$ vanishes and the vector current is conserved. The conservation
of the vector current implies that $h_+^{i\rightarrow i}(w=1)=1$. This condition, inserted
in the previous equations, translates into a condition on the derivatives of
$h_+^{i\rightarrow i}$ with respect to $\varepsilon_+$ 
\begin{eqnarray}
\frac{\partial^n h_+(w=1,0,0)}{\partial \varepsilon_+^n} = 0
\nonumber 
\end{eqnarray}
We thus expect that, for values of $w\simeq 1$ the corrections proportional 
to $\varepsilon_+$ will be rather small while the ones proportional to $\varepsilon_-$,
not constrained by the vector symmetry, can play a role also at zero recoil. 
This expectation is confirmed by our numerical results 
(see section~\ref{sec:fr}).
 
The discussion above is a re-derivation of the "Luke's theorem"~\cite{Luke:1990eg}. 
The theorem, originally derived by using HQET arguments, states that 
$h_+^{i\rightarrow i}$ it is not affected by first order corrections at zero recoil. 
In our language
\begin{eqnarray}
h_+(w=1,\varepsilon_+,\varepsilon_-) &=& 1 + 
\frac{\varepsilon_-^2}{2} \frac{\partial^2 h_+(w=1,0,0)}{\partial \varepsilon_-^2} +
\dots
\label{eq:taylorexp2}
\end{eqnarray}
Eqs.~(\ref{eq:taylorexp1}) confirm the analysis of the subleading corrections
to the form factors that has been carried out by the authors of ref.~\cite{Falk:1992wt}
within HQET\footnote{Some care is needed when eqs.~(\ref{eq:taylorexp1})
 are compared with the corresponding results of ref.~\cite{Falk:1992wt}. 
Indeed eqs.~(\ref{eq:taylorexp1}) are the result of a Taylor expansion and the coefficients 
do not depend upon $\varepsilon_{+,-}$ and, consequently, upon the
meson masses. 
Eqs.~(B1) of ref.~\cite{Falk:1992wt} are expansions in inverse powers of the quark masses
that depend upon the renormalization scale as well as the coefficients.
This dependence cancels at any given order.}.  
The additional symmetries of the static theory imply relations among the coefficients 
appearing in eqs.~(\ref{eq:taylorexp1}) and the corresponding ones arising in the 
case of vector-pseudoscalar and vector-vector transitions. In particular,
$h_+^{i\rightarrow i}$ is proportional at leading order to the Isgur-Wise 
function~\cite{Isgur:1989ed}.

In ref.~\cite{de Divitiis:2007ui} we have shown the results concerning the form factor
$G^{B\rightarrow D}(w)$ that enters into the semileptonic decay rate of a $B$ meson into a 
$D$ meson in the approximation of massless leptons $\ell=e,\mu$, 
\begin{eqnarray}
&&\frac{d\Gamma^{B\rightarrow D\ell\nu_{\ell}}}{dw}=
\vert V_{cb}\vert^2 \frac{G_F^2}{48\pi^3}(M_{B}+M_{D})^2M_{D}^3(w^2-1)^{3/2}
\left[ G^{B\rightarrow D}(w)\right]^2
\nonumber \\ \nonumber \\ \nonumber \\
&&1 \le w \le \frac{M_B^2+M_D^2}{2M_BM_D}
\nonumber 
\end{eqnarray}
This form factor is related to
$h_+^{i\rightarrow f}(w)$ and $h_-^{i\rightarrow f}(w)$ by
\begin{eqnarray}
G^{i\rightarrow f}(w)=h^{i\rightarrow f}_+(w)\ -\
\frac{M_f-M_i}{M_f+M_i}\ h^{i\rightarrow f}_-(w)
\nonumber
\end{eqnarray}
In the case $\ell=\tau$ the mass of the lepton cannot be neglected and the differential
decay rate is given by~\cite{Korner:1989qb,Kiers:1997zt}
\begin{eqnarray}
&&\frac{d\Gamma^{B\rightarrow D\tau\nu_{\tau}}}{dw}=
\frac{d\Gamma^{B\rightarrow D (e,\mu)\nu_{e,\mu}}}{dw}
\left(1-\frac{r_{\tau}^2}{t(w)} \right)^2
\left\{
\left(1+\frac{r_{\tau}^2}{2t(w)} \right)
+\frac{3r_{\tau}^2}{2t(w)}\frac{w+1}{w-1} \left[\Delta^{B\rightarrow D}(w)\right]^2
\right\}
\nonumber \\ \nonumber \\ \nonumber \\
&& r_{\tau}=\frac{m_{\tau}}{M_B}, \qquad r=\frac{M_D}{M_B}, \qquad t(w)=1+r^2-2rw,
\nonumber \\ \nonumber \\ \nonumber \\
&&1 \le w \le \frac{M_B^2+M_D^2-m_{\tau}^2}{2M_BM_D}
\nonumber 
\end{eqnarray}
In this work we provide an estimate of the function $\Delta^{B\rightarrow D}(w)$
appearing in the previous relations, including values at $w>1$. Its expression in terms of 
$h_+^{i\rightarrow f}(w)$ and $h_-^{i\rightarrow f}(w)$ is given by
\begin{eqnarray}
\Delta^{i\rightarrow f}(w) &=&
\frac{1}{G^{i\rightarrow f}(w)}
\left[\frac{1-r}{1+r}\ h^{i\rightarrow f}_+(w)\ -\ \frac{w-1}{w+1}\ h^{i\rightarrow f}_-(w) \right]
\nonumber \\ \nonumber \\ \nonumber \\
&=&\left(\frac{1-r}{1+r}\  -\ \frac{w-1}{w+1}\ \frac{h^{i\rightarrow f}_-(w)}{h^{i\rightarrow f}_+(w)}\right)
\left(1\ -\ \frac{1-r}{1+r}\ \frac{h^{i\rightarrow f}_-(w)}{h^{i\rightarrow f}_+(w)}\right)^{-1}
\label{eq:deltadef}
\end{eqnarray}
In the elastic case $\Delta^{i\rightarrow f}(w)$ vanishes identically and, in
the approximation in which $h^{i\rightarrow f}_-(w)$ is much smaller than $h^{i\rightarrow f}_+(w)$,
it is very well approximated by its static limit
\begin{eqnarray}
\Delta^{i\rightarrow f}(w) \simeq \frac{1-r}{1+r},\qquad\qquad r=\frac{M_f}{M_i}
\label{eq:deltastatic}
\end{eqnarray}
%

\section{Lattice Observables} \label{sec:notations}
We have carried out the calculation within the $O(a)$ improved 
Schr\"odinger Functional formalism~\cite{Luscher:1992an,Sint:1993un} with $T=2L$ and vanishing background fields. Physical units have been set by using the Sommer's scale 
and fixing $r_0=0.5$~fm~\cite{Guagnelli:1998ud,Necco:2001xg,Guagnelli:2002ia}.
In order to set the notations, we introduce the following source operators
\begin{eqnarray}
&&O_{sr}=\frac{a^6}{L^3}\sum_{{\bf y},{\bf z}}{\bar{\zeta}_s({\bf y})\gamma_5\zeta_r({\bf z})},
\quad
O^\prime_{sr}=\frac{a^6}{L^3}\sum_{{\bf y},{\bf z}}{\bar{\zeta}^\prime_s({\bf y})\gamma_5
\zeta_r^\prime({\bf z})}
\nonumber 
\end{eqnarray}
where $s$ and $r$ are flavour indexes while $\zeta$ and $\zeta^\prime$ are boundary fields at
$x_0=0$ and $x_0=T$ respectively.
The bulk operators are defined according to
\begin{eqnarray}
&&A^0_{sr}(x)=\bar{\psi}_s(x)\gamma_5\gamma^0\psi_r(x),
\qquad
P_{sr}(x)=\bar{\psi}_s(x)\gamma_5\psi_r(x)
\qquad
\mathcal{A}^0_{sr}(x)=A^0_{sr}(x)+ac_A\frac{\partial_0+\partial_0^*}{2}P_{sr}(x)
\nonumber \\ \nonumber \\
&&V^\mu_{sr}(x)=\bar{\psi}_s(x)\gamma^\mu\psi_r(x),
\quad
T^{\mu\nu}_{sr}(x)=\bar{\psi}_s(x)\gamma^\mu\gamma^\nu\psi_r(x)
\qquad
\mathcal{V}^\mu_{sr}(x)=V^\mu_{sr}(x)+ac_V\frac{\partial_\nu+\partial_\nu^*}{2}
T_{sr}^{\mu\nu}(x)
\nonumber
\end{eqnarray}
The improvement coefficient $c_A$ has been computed non-perturbatively
in ref.~\cite{Luscher:1996ug}. Regarding $c_V$, we have used the perturbative
result from ref.~\cite{Sint:1997jx} but its actual value influences our results
at the level of a few per mille.

The quark masses have been defined through the PCAC relation.
We have calculated the following correlation functions
\begin{eqnarray}
f^A_{sr}(x_0)=-\sum_{\bf x}{\langle O_{rs} A^0_{sr}(x)\rangle}
\qquad \qquad \qquad
f^P_{sr}(x_0)=-\sum_{\bf x}{\langle O_{rs} P_{sr}(x)\rangle}
\nonumber
\end{eqnarray}
and defined
\begin{eqnarray}
m_r^{AWI}=\frac{1}{2f^P_{rr}}\left[\frac{\partial_0+\partial_0^*}{2}f^A_{rr}+ac_A\partial_0\partial_0^*f^P_{rr}\right],
\qquad \qquad \qquad
a m_r^{VWI}=\frac{1}{2}\left[\frac{1}{k_r}-\frac{1}{k_c}\right]
\nonumber
\end{eqnarray}
where $a$ is the lattice spacing, $k_r$ is the hopping parameter of the $r$ quark and
$k_c$ is the critical value of the hopping parameter.
The renormalization group invariant (RGI) quark masses have been obtained by
the following relation
\begin{eqnarray}
m_r = Z_M\;\left[1+(b_A-b_P)\ am_r^{VWI} \right]\; m^{AWI}_r
\label{eq:rgi}
\end{eqnarray}
The combination $b_A-b_P$ of the improvement coefficients of the axial current and pseudoscalar density has been computed non-perturbatively 
in~\cite{deDivitiis:1997ka,Guagnelli:2000}. The factor $Z_M$ 
is known with very high precision in a range of inverse bare couplings that does not cover all 
the values of $\beta$ used in our simulations.
We have used the results reported in table~6 of ref.~\cite{Capitani:1998mq} 
to parametrize $Z_M$ in the enlarged range of $\beta$ values $[5.9,7.6]$.

In order to define on the lattice the matrix elements of the vector current between
pseudoscalar meson states, we need to introduce other two correlation functions,
\begin{eqnarray}
&&\mathcal{F}_{i\rightarrow f}^\mu(x_0;{\bf p_i},{\bf p_f})= \frac{a^3}{2}\sum_{\bf x}{
\langle O_{li} \ \mathcal{V}^\mu_{if}(x)\ O_{fl}^\prime \rangle
},
\qquad \qquad \qquad
f_{\mathcal{A}}^f(x_0,{\bf p_f})=-\sum_{\bf x}{
\langle O_{lf} \mathcal{A}^0_{fl}(x)\rangle}
\nonumber
\end{eqnarray}
where $i$ and $f$ are the heavy flavour indexes and $l$ is the light one.
The external momenta have been set by using flavour twisted b.c. for the 
heavy flavours; in particular we have used
\begin{eqnarray}
&&\psi_{i,f}(x+\hat{1}L)=e^{i\theta_{i,f}}\psi_{i,f}(x),
\qquad \qquad
p_1=\frac{\theta_{i,f}}{L}+\frac{2\pi k_1}{L},\qquad k_1\in \mathbb{N}
\nonumber
\label{eq:moms}
\end{eqnarray}
and ordinary periodic b.c. in the other spatial directions and for the light quarks. 
We have worked in the Lorentz frame in which the parent particle is at rest (${\bf p_i=0}$).
In this frame $w$ is simply expressed in terms of the ratio between the energy and the
mass of the daughter particle $w=E_f/M_f$.
The matrix elements of $V^\mu$ have been defined by the following ratios
\begin{eqnarray}
\langle V^\mu \rangle_{D1}^{i\rightarrow f}\equiv
\langle \mathcal{M}_f\vert \ V^\mu \ \vert \mathcal{M}_i\rangle_{D1}\equiv
2\sqrt{M_i E_f}\frac{\mathcal{F}_{i\rightarrow f}^\mu(T/2;{\bf 0},{\bf p_f})}{
\sqrt{
\mathcal{F}_{i\rightarrow i}^0(T/2;{\bf 0},{\bf 0})
\mathcal{F}_{f\rightarrow f}^0(T/2;{\bf p_f},{\bf p_f})
}}
\label{eq:def1}
\end{eqnarray}
that become the physical matrix elements in large volumes where single state
dominance is a good approximation.
An alternative definition of the matrix elements ($D2$), which 
reduces to the previous one ($D1$) in the infinite volume and at zero lattice spacing, 
can be obtained by considering
\begin{eqnarray}
\langle V^\mu \rangle_{D2}^{i\rightarrow f}\equiv
\langle \mathcal{M}_f\vert \ V^\mu \ \vert \mathcal{M}_i\rangle_{D2}\equiv
2\frac{\sqrt{M_i} E_f f_{\mathcal{A}}^f(T/2,{\bf 0})}{\sqrt{M_f}f_{\mathcal{A}}^f(T/2,{\bf p_f})}
\frac{\mathcal{F}_{i\rightarrow f}^\mu(T/2;{\bf 0},{\bf p_f})}{
\sqrt{
\mathcal{F}_{i\rightarrow i}^0(T/2;{\bf 0},{\bf 0})
\mathcal{F}_{f\rightarrow f}^0(T/2;{\bf 0},{\bf 0})
}}
\label{eq:def2}
\end{eqnarray}
In eqs.~(\ref{eq:def1}) and~(\ref{eq:def2}) the renormalization factors $Z_V$ and $Z_A$
cancel in the ratios together with the factors containing the improvement coefficients
$b_V$ and $b_A$.

By calculating the following ratio
\begin{eqnarray}
x_f\;=\;\frac{\mathcal{F}_{f\rightarrow f}^1(T/2;{\bf 0},{\bf p_f})}
{\mathcal{F}_{f\rightarrow f}^0(T/2;{\bf 0},{\bf p_f})}
\;=\; \frac{\langle \mathcal{M}_f\vert \ \mathcal{V}^1 \ \vert \mathcal{M}_f\rangle}
{\langle \mathcal{M}_f\vert \ \mathcal{V}^0 \ \vert \mathcal{M}_f\rangle}
\;=\; \frac{\sqrt{w^2-1}}{w+1} \nonumber
\end{eqnarray}
we have defined $w$, as well as  meson masses and energies, entirely in terms of three point
correlation functions. This definition of $w$ is noisier than the one that can be obtained
in terms of ratios of two point correlation functions; however it leads to exact vector
current conservation when $M_f=M_i$ and reduces the final statistical error on the
form factors. The two definitions of the matrix elements lead to two definitions of the form factors that, in terms of $\langle V^0 \rangle_D$ and $\langle V^1 \rangle_D$, are
expressed by
\begin{eqnarray}
&&h_+^{i\rightarrow f}(w) \;=\; \frac{\langle V^0 \rangle^{i\rightarrow f}}{2M_i\sqrt{r}}\;
\left\{
1\;
+ \;\frac{\sqrt{w^2-1}}{w+1}
\;
\frac{\langle V^1 \rangle^{i\rightarrow f}}
{\langle V^0 \rangle^{i\rightarrow f}}\right\}
\label{eq:hpdefme}
\\ \nonumber \\ \nonumber \\
&&h_-^{i\rightarrow f}(w) \;=\; \frac{\langle V^0 \rangle^{i\rightarrow f}}{2M_i\sqrt{r}}
\left\{
1\;
+ \;\frac{w+1}{\sqrt{w^2-1}}
\;
\frac{\langle V^1 \rangle^{i\rightarrow f}}
{\langle V^0 \rangle^{i\rightarrow f}}\right\}
\label{eq:hmdefme}
\\ \nonumber \\ \nonumber \\
&&G^{i\rightarrow f}(w) \;=\; \frac{2r}{1+r}\;
\frac{\langle V^0 \rangle^{i\rightarrow f}}{2M_i\sqrt{r}}
\left\{1\;
+ \;\frac{wr-1}{r\sqrt{w^2-1}}
\;
\frac{\langle V^1 \rangle^{i\rightarrow f}}
{\langle V^0 \rangle^{i\rightarrow f}}\right\},
\qquad \qquad
r=\frac{M_f}{M_i}
\label{eq:gdefme}
\end{eqnarray}
The last two equations are not defined at $w=1$; this is due to the second term
in the parenthesis of eq.~(\ref{eq:hmdefme}) and~(\ref{eq:gdefme})
that we extrapolate at zero recoil before calculating $h_-^{i\rightarrow f}(w=1)$
and $G^{i\rightarrow f}(w=1)$.
\begin{figure}[t]
\includegraphics[width=0.6\textwidth]{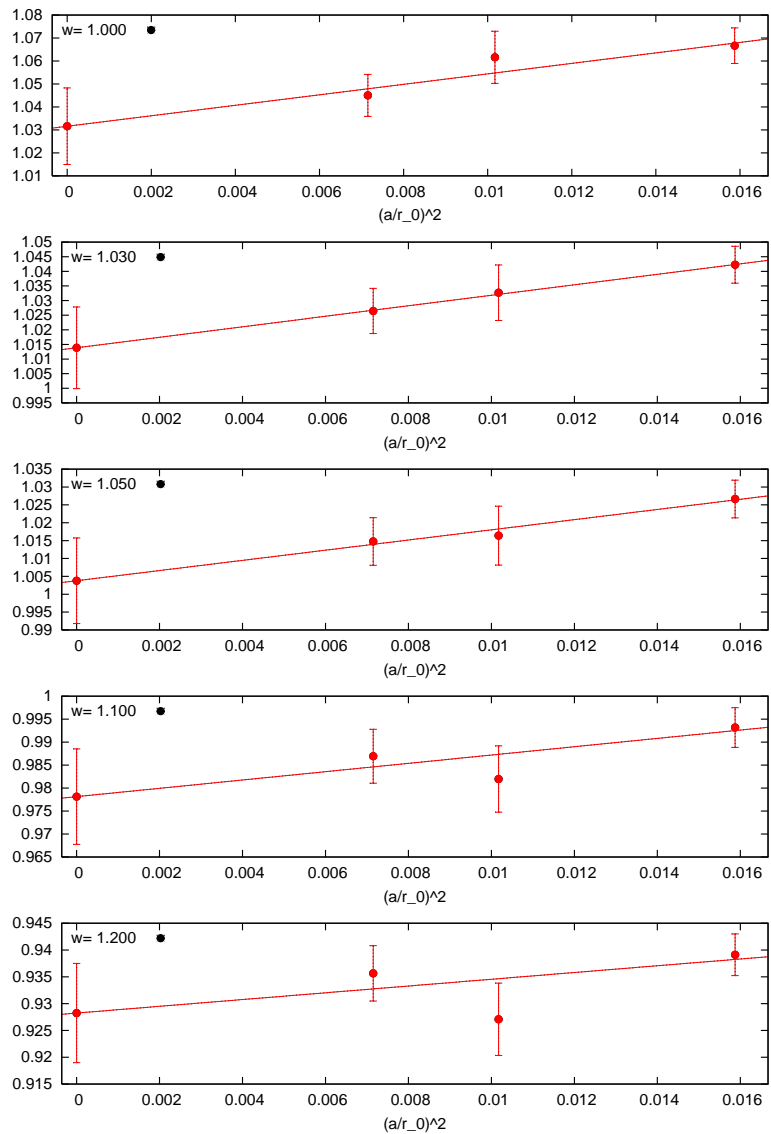}
\caption{\label{fig:climits0} Extrapolations to the continuum
limit of $G^{B\rightarrow D}(w)$. The data correspond to $m_l=m_s$, to
the definition $D1$ and to the data sets $L_0A$, $L_0B$ and $L_0C$.} 
\end{figure}
%

\section{the step scaling method} \label{sec:ssm}
The SSM has been introduced to cope with two-scale problems in lattice QCD. In the
calculation of heavy-light meson properties the two scales are the mass
of the heavy quarks ($b$,$c$) and the mass of the light quarks ($u$,$d$,$s$). 
Here we consider the generic form factor 
$F^{i\rightarrow f}=\{h_+^{i\rightarrow f},h_-^{i\rightarrow f},G^{i\rightarrow f}\}$
as a function of $w$, the volume $L^3$ and fix the meson states by the corresponding
heavy and light RGI quark masses that, being extracted by the lattice version of the 
PCAC relation, are not affected by finite volume effects (see eq.~\ref{eq:rgi}). 

The first step of the finite volume recursion consists in calculating
the observable $F^{i\rightarrow f}(w;L_0)$ on a small volume, $L_0$, 
which is chosen to accommodate the dynamics of heavy quarks with masses ranging from the 
physical value of the charm mass up to the mass of the bottom. 
As in our previous work we fixed $L_0=0.4$~fm.
We have simulated five different heavy quark masses 
$m_{i,f}=\{m_h^1,m_h^2,m_h^3,m_h^4,m_h^5\}$, five different momenta
$\theta_1=\{\theta^1_1,\theta^2_1,\theta^3_1,\theta^4_1,\theta^5_1\}$ 
(see eq.~(\ref{eq:moms}))  and three light quark masses $m_l=\{m_l^1,m_l^2,m_l^3\}$.  

A first effect of finite volume is taken into account by evolving
the results from $L_0$ to $L_1=0.8$~fm through the factor
\begin{displaymath} 
\sigma^{i\rightarrow f}(w;L_0,L_1)=\frac{F^{i\rightarrow f}(w;L_1)}{F^{i\rightarrow f}(w;L_0)}
\end{displaymath}
computed for each value of $w$ and for each value of the light quark mass.
The crucial point is that the step scaling functions are calculated by simulating
heavy quark masses smaller than the $b$-quark mass. 
The step scaling functions at $m_i\simeq m_b$ and $m_f\simeq m_c$
are obtained by directly simulating $m_f$ both on $L_0$ and on $L_1$ and
by a smooth extrapolation in $1/m_i$. 

Extrapolating the step scaling functions is more advantageous than extrapolating 
the form factors. This can be easily understood by relying on HQET expectations 
(see also eq.~(\ref{eq:taylorexp1})),
\begin{eqnarray} 
\sigma^{i\rightarrow f}(w;L_0,L_1) &=&
\frac{
F^{(0) \rightarrow f}(w;L_1)\;\left[1+
\frac{F^{(1) \rightarrow f}(w;L_1)}{m_i}+\dots\right]}
{F^{(0) \rightarrow f}(w;L_0)\;\left[1+
\frac{F^{(1) \rightarrow f}(w;L_0)}{m_i}+\dots\right]}
\nonumber \\ \nonumber \\ \nonumber \\
&=&
\frac{F^{(0) \rightarrow f}(w;L_1)}{F^{(0) \rightarrow f}(w;L_0)}
\;\left[
1+\frac{F^{(1) \rightarrow f}(w;L_1)-F^{(1) \rightarrow f}(w;L_0)}{m_i}
+\dots\right]
\nonumber \\ \nonumber \\ \nonumber \\
&\equiv&
\sigma^{(0)\rightarrow f}(w;L_0,L_1)
\;\left[
1+\frac{\sigma^{(1)\rightarrow f}(w;L_0,L_1)}{m_i}
+\dots\right]
\nonumber 
\end{eqnarray}
In the previous relations the superscripts in parenthesis, $(n)$, 
mark the order of the expansion in the inverse heavy quark mass.
The subleading correction to the step scaling functions is the difference
of two terms and vanishes in the infinite volume,
$\sigma^{(1)\rightarrow f}(w;L_0,L_1)=F^{(1) \rightarrow f}(w;L_1)-F^{(1) \rightarrow f}(w;L_0)$, becoming smaller and smaller as the volume is increased.
This matches the general idea that finite volume effects, measured
by the $\sigma$'s, are almost insensitive to the high energy scale.

We also compute the step scaling functions of the elastic form factors 
$h_+^{i\rightarrow i}$ at $m_i\simeq m_b$ by extrapolating the corresponding
results from smaller heavy quark masses. Also in this case
the $\sigma$'s are expected to be almost flat with respect to $1/m_i$.

In order to remove the residual finite volume effects we iterate the procedure
described above once more passing from $L_1$ to $L_2=1.2$~fm. Our final results
are obtained from
\begin{eqnarray} 
F^{i\rightarrow f}(w;L_2) \quad=\quad
F^{i\rightarrow f}(w;L_0)\quad
\sigma^{i\rightarrow f}(w;L_0,L_1)\quad
\sigma^{i\rightarrow f}(w;L_1,L_2)
\label{eq:ssm}
\end{eqnarray}
\begin{figure}[t]
\includegraphics[width=\textwidth]{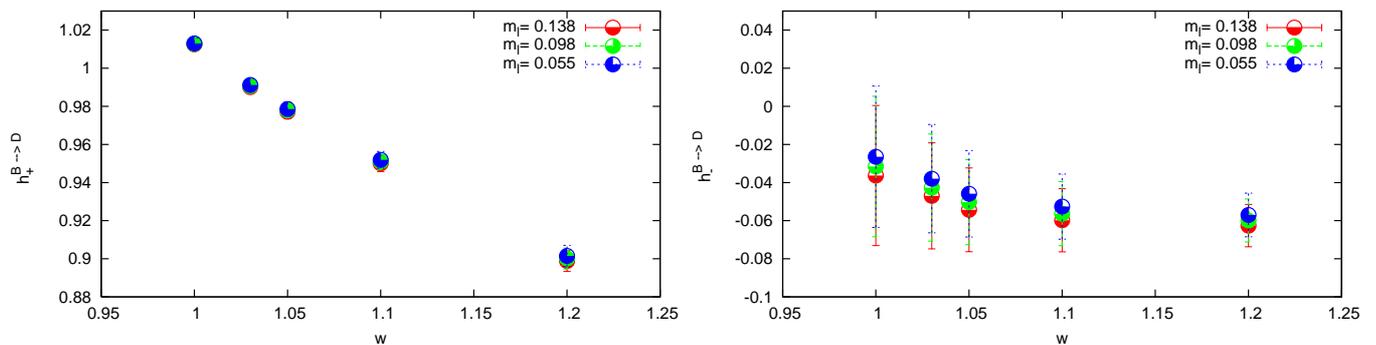}
\caption{\label{fig:chirals0} Light quark mass dependence of 
$h_+^{B\rightarrow D}(w;L_0)$ (left) and of $h_-^{B\rightarrow D}(w;L_0)$ (right).
The different sets of points correspond to different values of $m_l$ ranging from
about $m_s$ to about $m_s/4$. The data are in the continuum limit
and correspond to the definition $D1$.} 
\end{figure}
\begin{figure}[t]
\includegraphics[width=\textwidth]{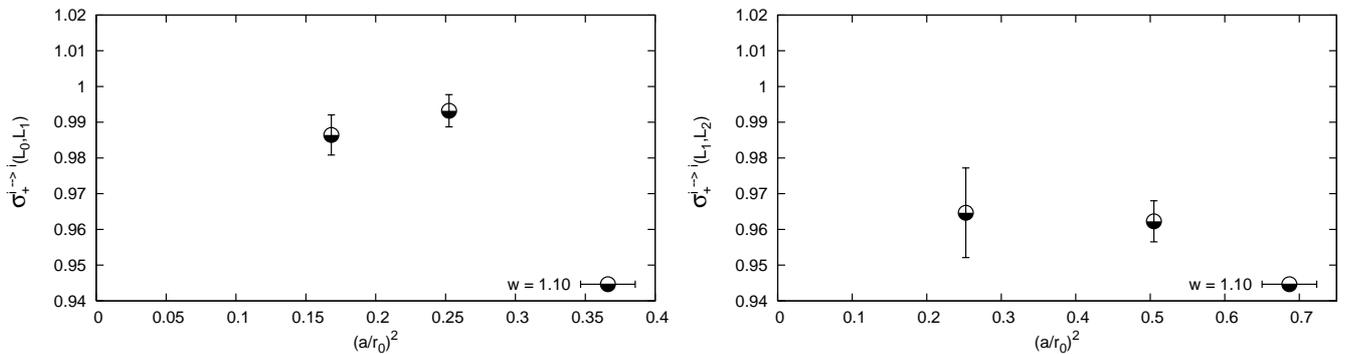}
\caption{\label{fig:ssfcl} Continuum extrapolation of $\sigma_+^{i\rightarrow i}(w=1.1;L_0,L_1)$ (left)
and $\sigma_+^{i\rightarrow i}(w=1.1;L_1,L_2)$ (right) at the heaviest values of the
heavy quark masses ($m_i\simeq m_b/4$ and $m_i\simeq m_b/2$ respectively).
The data correspond to $m_l=m_s$, to the definition $D1$ and
to the data sets $L_1A/L_0a$, $L_1B/L_0b$ (left) and $L_2A/L_1a$, $L_2B/L_1b$ (right).} 
\end{figure}
\begin{figure}[t]
\includegraphics[width=\textwidth]{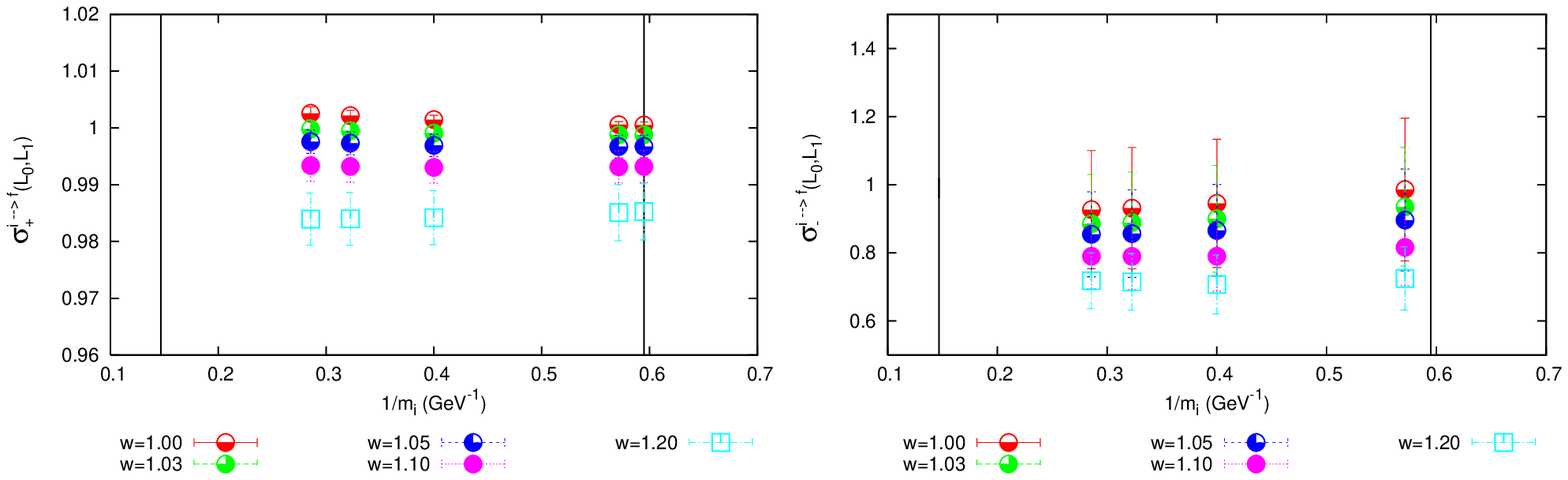}
\caption{\label{fig:ssf1} Step scaling functions  of $h_+^{i\rightarrow c}$ (left)
and $h_-^{i\rightarrow c}$ (right) as functions of $1/m_i$ for the first
evolution step (from $L_0$ to $L_1$). The black vertical
lines represent the physical points $m_i=m_c$ and $m_i=m_b$.
The data are in the continuum and chiral limits and correspond to the definition $D1$.} 
\end{figure}
\begin{figure}[t]
\includegraphics[width=\textwidth]{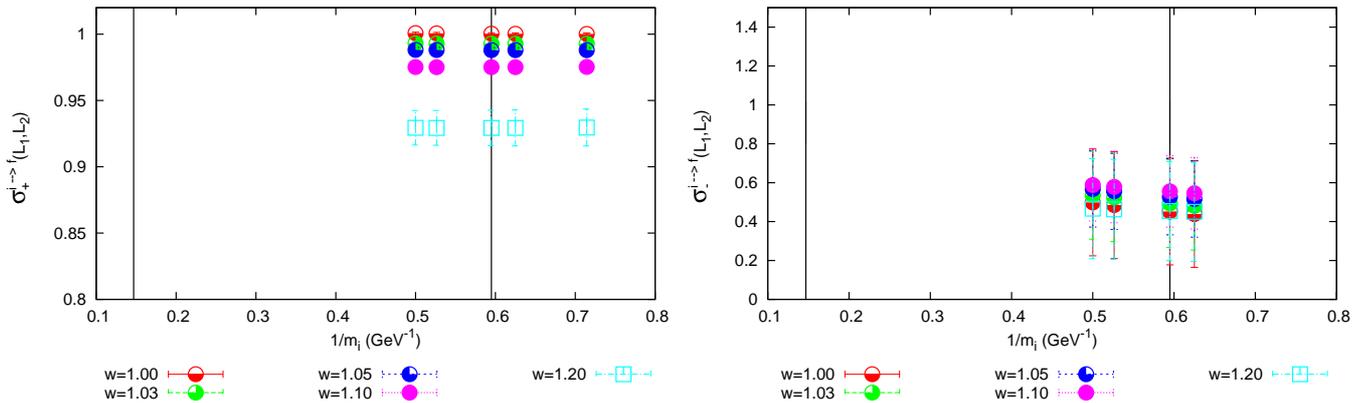}
\caption{\label{fig:ssf2} Step scaling functions  of $h_+^{i\rightarrow f}$ (left)
and $h_-^{i\rightarrow f}$ (right) at fixed $m_f$ as functions of $1/m_i$ for the second
evolution step (from $L_1$ to $L_2$). The black vertical
lines represent the physical points $m_i=m_c$ and $m_i=m_b$.
The data are in the continuum and chiral limits and correspond to the definition $D1$.} 
\end{figure}
\begin{figure}[t]
\includegraphics[width=\textwidth]{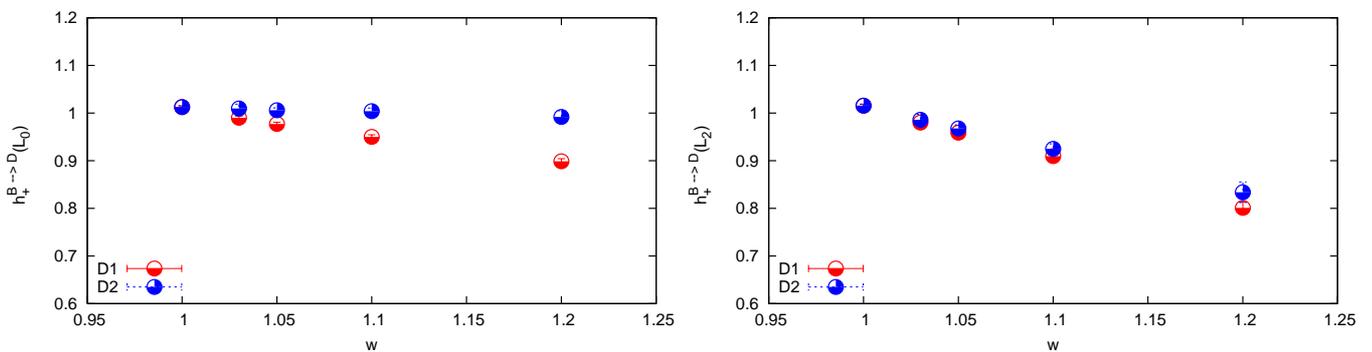}
\caption{\label{fig:2defs} Comparison of the two definitions of $h_+^{B\rightarrow D}(w;L)$
at $L_0=0.4$~fm (left) and at $L_2=1.2$~fm (right).
The data are in the continuum and chiral limits.} 
\end{figure}
\begin{figure}[t]
\includegraphics[width=\textwidth]{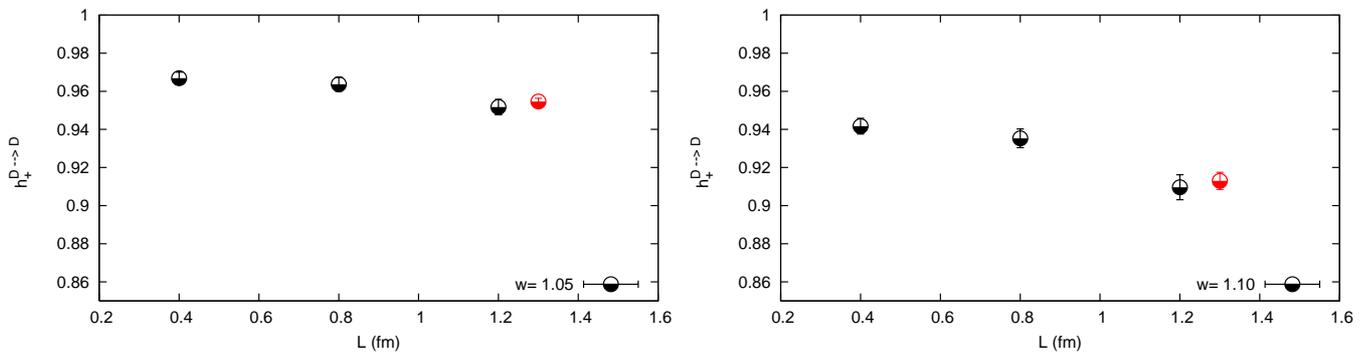}
\caption{\label{fig:voldep} $h_+^{D\rightarrow D}(w=1.05;L)$ (left)
and $h_+^{D\rightarrow D}(w=1.10;L)$ (right)
as functions of the volume.
The black points have been obtained through the step scaling recursion
while the red points (slightly displaced on the $x$-axis to help the eye)
are the result of a direct simulation on the biggest volume.
The data are in the continuum and chiral limits and correspond to the definition $D1$.} 
\end{figure}
\begin{figure}[t]
\includegraphics[width=\textwidth]{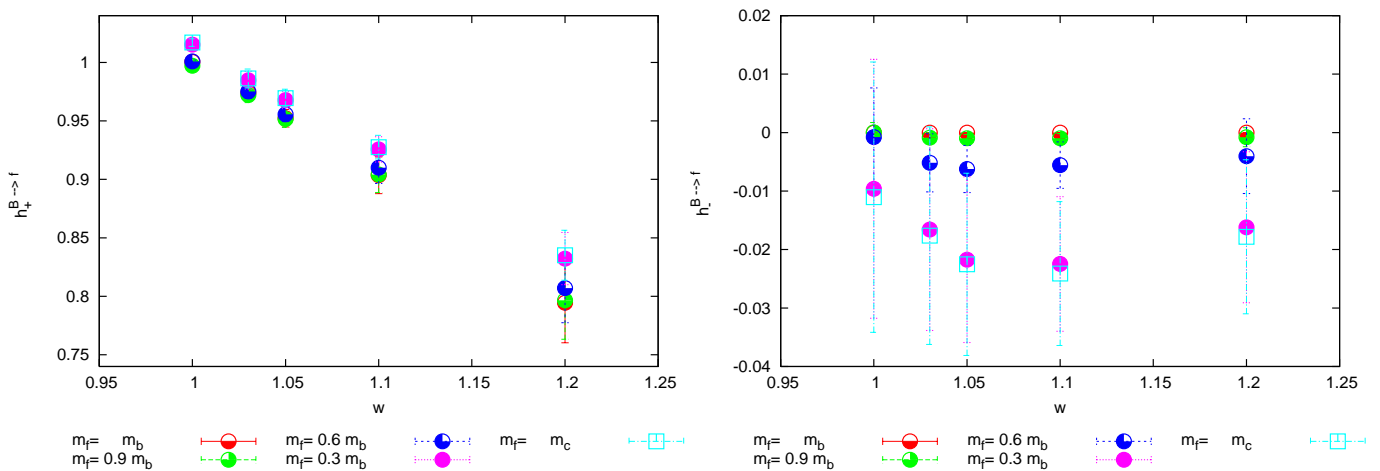}
\caption{\label{fig:iw1} 
In the left plot it is shown $h_+^{B\rightarrow f}(w)$ in the
infinite volume limit as a function of $w$ for different values of the 
final heavy quark mass. The right plot shows $h_-^{B\rightarrow f}(w)$
for the same combinations of heavy quark masses.} 
\end{figure}
\begin{figure}[t]
\includegraphics[width=\textwidth]{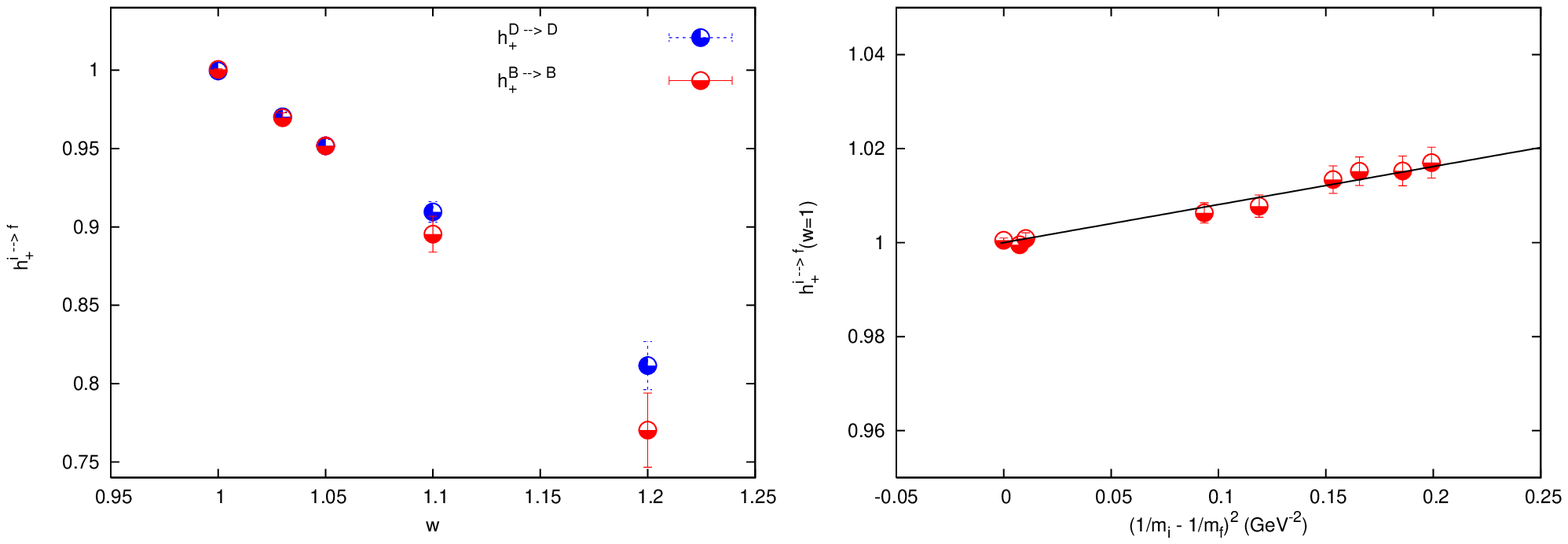}
\caption{\label{fig:iw2} 
The left plot shows $h_+^{B\rightarrow B}(w)$
and $h_+^{D\rightarrow D}(w)$: in the range $1\le w \le 1.05$ 
the two elastic form factors are indistinguishable within the quoted errors while 
$h_+^{B\rightarrow D}(w)$ in fig.~\ref{fig:iw1} shows appreciable corrections from the 
Isgur-Wise limit, in particular at zero recoil.
The right plot shows $h_+^{i\rightarrow f}$ at zero recoil ($w=1$) as a function
of $\varepsilon_-^2$ (actually $(1/m_i-1/m_f)^2 \propto \varepsilon_-^2$,
$m_{i,f}$ being the RGI heavy quark masses): the solid line has been obtained
by fitting the data according to eq.~(\ref{eq:taylorexp2}). } 
\end{figure}
%

\section{finite volume results} \label{sec:fvr}

\subsection{small volume}
The small volume $L_0=0.4$~fm has been simulated by using three different values of the 
lattice spacing (see table~\ref{tab:sims0}). The small physical extent of the volume 
allowed us to simulate
relativistic heavy quarks with masses ranging from around $m_b$ down to $m_c$.
We have computed the form factors $h_+^{i\rightarrow f}$, $h_-^{i\rightarrow f}$
and $G^{i\rightarrow f}$ for all the combinations of heavy and light quark masses and
for five different values of the momentum transfer. 

In figure~\ref{fig:climits0} we show the continuum extrapolations of $G^{B\rightarrow D}(w)$. 
The points in this figure correspond to $m_l=m_s$ but similar figures can be obtained for the 
other values of the light and heavy quark masses and for the other form factors.

In figure~\ref{fig:chirals0} we show $h_+^{B\rightarrow D}$ (left) and
$h_-^{B\rightarrow D}$ (right) as functions of $w$ for the three 
different values of light quark masses that we have simulated (ranging from about $m_s$
to $m_s/4$). 
As we have anticipated in ref.~\cite{de Divitiis:2007ui}, we find that
the $F$'s behave as constants with respect to $m_l$ within the statistical errors.
This happens for each combination of heavy quark masses and for each value of the lattice 
spacing. Nevertheless we make a linear extrapolation to reach the chiral limit;
the resulting error largely accounts for the systematics due to these extrapolations.
In the following our results include the mild chiral extrapolation.

\subsection{steps}
The parameters of the simulations of the evolution steps are given
in tables~\ref{tab:sims1} and~\ref{tab:sims2} . 
We have been simulating at two different lattice spacings by limiting
the maximum value of the heavy quark mass to $m_i\simeq m_b/2$ for
the first step and to $m_i\simeq m_b/4$ for the second. 
In figure~\ref{fig:ssfcl} we show the dependence upon the lattice spacing of
the step scaling functions $\sigma_+^{i\rightarrow f}(w=1.1;L_0,L_1)$ (left)
and $\sigma_+^{i\rightarrow f}(w=1.1;L_1,L_2)$ (right) in the worst case (largest value
of heavy quark masses). 
In general our results are consistent
with a scaling regime within a few per mille accuracy and the continuum 
step scaling functions of tables~\ref{tab:s1results} and~\ref{tab:s2results}
have been obtained by averaging the results at the two lattice spacings.

In figure~\ref{fig:ssf1} we can test our hypothesis on
the low sensitivity of the step scaling functions upon the high energy scale.
The figure shows the step scaling functions of the form factors $h_+^{i\rightarrow c}$ (left)
and $h_-^{i\rightarrow c}$ (right) as functions of $1/m_i$. In both cases the dependence
upon $m_i$ is hardly appreciable and in the case of $h_+^{i\rightarrow c}$ the $\sigma$'s are 
very close to one while $h_-^{i\rightarrow c}$ is affected by stronger finite volume effects. 
We obtain the values at $m_i=m_b$ by linear fits. 

In figure~\ref{fig:ssf2} we plot the same quantities as in figure~\ref{fig:ssf1} for
the second evolution step (from $L_1$ to $L_2$, see table~\ref{tab:sims2}). Also
in this case the step scaling functions depend very smoothly upon $1/m_i$.

\subsection{consistency checks}
In this section we illustrate the results of two checks that we have done in order to
convince ourselves on the consistency of the step scaling procedure.
As already discussed in sec.~\ref{sec:notations}, we have used two different definitions
of the matrix elements and, consequently, of each form factor. In figure~\ref{fig:2defs}
we show the comparison of $h_+^{B\rightarrow D}(w;L)$ at $L_0=0.4$~fm (left) and at 
$L_2=1.2$~fm (right). We see that the results, while differing at finite volume, 
converge to common values after the step scaling procedure. 
This makes us confident of a correct accounting of finite volume effects.

A second check of the whole procedure, and in particular of the continuum limit of the
step scaling functions, can be obtained by considering the elastic
form factor $h_+^{D\rightarrow D}(w;L)$ at fixed $w$ as a function of $L$.
The point is that the charm quark mass has been simulated directly on each physical
volume and, in particular, on the biggest one. In figure~\ref{fig:voldep} we fix 
$w=1.05$ (left) and $w=1.10$ (right) and 
see that the step scaling recursion (black points) converge to the result obtained
directly at $L_2=1.2$~fm (red points, slightly displaced to help the eye) making
us confident of a correct accounting of the cutoff effects and, in particular,
of a correct estimate of the error on the continuum step scaling functions. 

Our final results are obtained by averaging over the two definitions
and by combining in quadrature statistical errors with the systematic ones 
that we estimate from the dispersion between $D1$ and $D2$.

\section{final results} \label{sec:fr}
In this section we discuss our final results in the continuum, chiral and infinite volume
limits (table~\ref{tab:endofthestory}). 
In order to establish the onset of
the static limit approximation we plot in figure~\ref{fig:iw1} the form factor $h_+^{B\rightarrow f}(w)$ 
as a function of $w$ for different values of the 
final heavy quark mass. The right plot shows $h_-^{B\rightarrow f}(w)$
for the same combinations of heavy quark masses.
We see that the corrections to the static limits of both $h_+^{B\rightarrow f}(w)$ and
$h_-^{B\rightarrow f}(w)$ are of the order of $2$\% at the charm mass.
For heavy quark masses bigger than $m_b/2$ the corrections are almost negligible
(below $1$\%).

Eqs.~(\ref{eq:taylorexp1}) and~(\ref{eq:taylorexp2}) predict that the convergence toward
the static limit is faster in the case of the elastic form factors with respect to the
ones having $m_i>m_f$. This happens because near the point at zero recoil the subleading
corrections to $h_+^{i\rightarrow f}(w)$ are proportional to the square of the difference
of the initial and final meson masses.
Figure~\ref{fig:iw2} clearly shows that this happens in practice. Indeed, in the left plot we
see that the elastic form factor $h_+^{D\rightarrow D}(w)$ is much closer to
the static limit (very well approximated by $h_+^{B\rightarrow B}(w)$) with respect to
the form factor $h_+^{B\rightarrow D}(w)$, 
the one relevant into the calculation of $V_{cb}$ shown in figure~\ref{fig:iw1}. 
In the right plot of figure~\ref{fig:iw2} we show how well 
eq.~(\ref{eq:taylorexp2}) is approximated by our numerical data. The fit is performed on the
slope while the intercept is fixed to one. 

The QCD form factor $h_+^{i\rightarrow f}(w)$ is related to the renormalization 
group invariant HQET Isgur-Wise function, $\xi(w)$, by the 
following relation~\cite{Neubert:1992tg,Neubert:1992hb}
\begin{eqnarray}
h_+^{i\rightarrow f}(w) = \left[1+\beta_+(m_i,m_f;w)+\gamma_+(m_i,m_f;w) + O(m_{i,f}^{-2})\right]\xi(w)
\nonumber
\end{eqnarray}
where the $\gamma_+$ term accounts for non-perturbative power corrections proportional
to the inverse of the quark masses while the $\beta_+$ term accounts for perturbative
radiative corrections.
In the case of the elastic form factor $h_+^{i\rightarrow i}(w)$ at the highest value of 
the simulated heavy quark masses, i.e. the bottom quark mass,
power corrections are completely negligible in our data 
as clearly emerges from figures~\ref{fig:iw1} and~\ref{fig:iw2}:
\begin{eqnarray}
&&\gamma_+(m_b,m_b;w)\cong 0
\nonumber \\ \nonumber\\
&&h_+^{B\rightarrow B}(w) = \left[1+\beta_+(m_b,m_b;w) \right]\xi(w)
\nonumber
\end{eqnarray}
The function $\beta_+(m_b,m_b;w)$ depends logarithmically upon the bottom mass 
through $\alpha_s(m_b)$ and vanishes at zero recoil
where the renormalized Isgur-Wise function is identically equal to one like
the relativistic QCD elastic form factor. These terms are of the percent order
and their logarithmic dependence upon $m_b$ cannot be extrapolated away from
our data. Nevertheless, in order to get the HQET Isgur-Wise function
our non-perturbative results for $h_+^{B\rightarrow B}(w)$ 
(given in table~\ref{tab:endofthestory})
can be further corrected by hand through the perturbative $\beta_+(m_b,m_b;w)$
given in ref.~\cite{Neubert:1992tg} at next to leading order\footnote{for a recent
lattice calculation of the Isgur-Wise function see ref.~\cite{Bowler:2002zh}}.
\begin{figure}[t]
\includegraphics[width=0.8\textwidth]{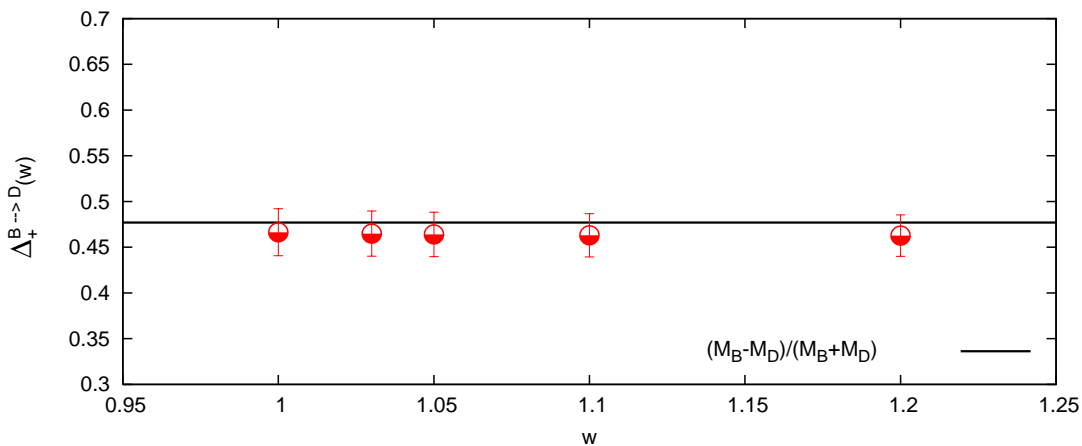}
\caption{\label{fig:delta} The figure shows the function $\Delta^{B\rightarrow D}(w)$
in the chiral, continuum, and infinite volume limits. The solid line correspond to the static limit
result, $(M_B-M_D)/(M_B+M_D)$, and has been drawn by using the experimental determinations
of the meson masses.} 
\end{figure}

Finally, we show in figure~\ref{fig:delta} our best result for the function $\Delta^{B\rightarrow D}(w)$
that enters in the decay rate of the process $B\rightarrow D\tau\nu_{\tau}$ (see discussion
at the end of section~\ref{sec:ff}).
$\Delta^{B\rightarrow D}(w)$ does not show any significant dependence upon
$w$ and is very well approximated by its static limit (see eq.~\ref{eq:deltastatic}).
These findings represent a prediction that can be confirmed by a future measurement of the 
differential decay rate of the process $B\rightarrow D\tau\nu_{\tau}$. 
Indeed, the function $\Delta^{B\rightarrow D}(w)$
can be extracted experimentally by the ratio 
$d\Gamma^{B\rightarrow D\tau\nu_{\tau}}/d\Gamma^{B\rightarrow D (e,\mu)\nu_{e,\mu}}$ that does not
depend upon the CKM matrix element. On the other hand, the knowledge of $\Delta^{B\rightarrow D}(w)$
is required in order to perform lepton-flavour universality checks on the extraction
of $V_{cb}$.

\section{conclusions} \label{sec:conc}
We have performed the calculation of the form factors that parametrize semileptonic
transitions among pseudoscalar heavy-light mesons and made a prediction for the ratio 
$d\Gamma^{B\rightarrow D\tau\nu_{\tau}}/d\Gamma^{B\rightarrow D (e,\mu)\nu_{e,\mu}}$.
In view of a future measurement of the differential
decay rate of the process $B\rightarrow D \tau \nu_\tau$, our results
will allow to perform lepton-flavour universality checks on the
extraction of $V_{cb}$.

The form factors have
been obtained with a relative accuracy of the order of a few percent allowing to
establish the range of validity of the heavy quark effective theory for
these quantities. 
In particular we have obtained a check of the predictions of the 
Luke's theorem that we re-derived. 
The corrections to the static limit are very small already in the case of the elastic form 
factor $h_+^{D\rightarrow D}$ and negligible in the case of $h_+^{B\rightarrow B}$.
We have also established the accuracy of the static approximation to the form factors
of the decay $B\rightarrow D\ell\nu$ which is of the order of $2$-$3$\% at zero recoil
and reaches about $7$\% at $w=1.2$ where becomes definitely inadequate for precise
phenomenological applications.
\begin{table}[t]
\begin{ruledtabular}
\begin{tabular}{cccccc}
$i\rightarrow f$ & $w$ & $G$ & $h_+$ & $h_-$ & $\Delta$ \\ 
\hline
                           & 1.00          & 1.000(00)  & 1.000(00) &            & \\
                           & 1.03          & 0.971(07)  & 0.971(07) &            & \\
$D\rightarrow D$           & 1.05          & 0.955(06)  & 0.955(06) &            & \\
                           & 1.10          & 0.916(09)  & 0.916(09) &            & \\
                           & 1.20          & 0.828(20)  & 0.828(20) &            & \\
\hline
                           & 1.00          & 1.000(00)  & 1.000(00) &            & \\
                           & 1.03          & 0.974(07)  & 0.974(07) &            & \\
$B\rightarrow B$           & 1.05          & 0.952(07)  & 0.952(07) &            & \\
                           & 1.10          & 0.903(16)  & 0.903(16) &            & \\
                           & 1.20          & 0.794(34)  & 0.794(34) &            & \\
\hline
                           & 1.00          & 1.026(17)  & 1.017(03) & -0.011(23) & 0.466(26)\\
                           & 1.03          & 1.001(19)  & 0.986(08) & -0.018(19) & 0.465(25)\\
$B\rightarrow D$           & 1.05          & 0.987(15)  & 0.970(07) & -0.023(16) & 0.464(24)\\
                           & 1.10          & 0.943(11)  & 0.928(10) & -0.024(12) & 0.463(24)\\
                           & 1.20          & 0.853(21)  & 0.835(21) & -0.018(13) & 0.463(23)\\
\end{tabular}
\end{ruledtabular}
\caption{\label{tab:endofthestory}
Physical results. Average of the two definitions $D1$ and $D2$.}
\end{table}

Our results have been obtained within the quenched approximation and further calculations
will be needed to asses the corrections due to unquenching.
On the other hand, the accuracy reached in the quenched case demonstrates
the feasibility and the opportunity of repeating the present calculation in the
unquenched theory. Indeed, the recursive matching process can be
extended to the sea quark masses that, alternatively, can be kept to their physical values
if the Schr\"odinger Functional formalism is used.
Moreover, flavour twisted boundary conditions can be used for heavy valence quarks 
also in the $N_f=3$ unquenched theory. The real case will further differ by the
heavy flavour determinants that can be accounted for by a perturbative expansion
in the hopping parameter.

\begin{acknowledgments}
We warmly thank E.~Molinaro for his participation at an early stage of this work.
The simulations required to carry on this project 
have been performed on the INFN apeNEXT machines at Rome "La Sapienza".
We thank A.~Lonardo, D.~Rossetti and P.~Vicini for technical advice. 
\end{acknowledgments}



%
\begin{table}
\begin{ruledtabular}
\begin{tabular}{lccccc}
 & $\beta$ & $T\times L^3$ & $N_{cnfg}$ & $k$ & $\theta$\\
\hline
$L_0A$ & 7.300 & $48\times 24^3$   & 277  & 0.124176 & 0.000000\\
       &       &                   &      & 0.124844 & 0.953456\\
       &       &                   &      & 0.128440 & 1.201080\\
       &       &                   &      & 0.129224 & 2.042983\\
       &       &                   &      & 0.131950 & 2.573569\\
       &       &                   &      & 0.134041 & \\
       &       &                   &      & 0.134098 & \\
       &       &                   &      & 0.134155 & \\
\hline
$L_0B$ & 7.151 & $40\times 20^3$   & 224 & 0.122666 & 0.000000\\
       &       &                   &     & 0.123437 & 0.953456\\
       &       &                   &     & 0.127605 & 1.201079\\
       &       &                   &     & 0.131511 & 1.719170\\
       &       &                   &     & 0.131686 & 2.488490\\
       &       &                   &     & 0.134277 & \\
       &       &                   &     & 0.134350 & \\
       &       &                   &     & 0.134422 & \\
\hline
$L_0C$ & 6.963 & $32\times 16^3$   & 403 & 0.120081 & 0.000000\\
       &       &                   &     & 0.120988 & 0.953456\\
       &       &                   &     & 0.126050 & 1.201079\\
       &       &                   &     & 0.131082 & 1.719170\\
       &       &                   &     & 0.131314 & 2.488490\\
       &       &                   &     & 0.134526 & \\
       &       &                   &     & 0.134614 & \\
       &       &                   &     & 0.134702 & \\
\end{tabular}
\end{ruledtabular}
\caption{\label{tab:sims0}
Table of lattice simulations of the small volume.}
\end{table}
\begin{table}
\begin{ruledtabular}
\begin{tabular}{ccccc}
$i\rightarrow f$ & $w$ & $G$ & $h_+$ & $h_-$ \\ 
\hline
                           & 1.00          & 1.000(00) & 1.000(00) & \\
                           & 1.03          & 0.979(02) & 0.979(02) & \\
$D\rightarrow D$           & 1.05          & 0.967(03) & 0.967(03) & \\
                           & 1.10          & 0.942(04) & 0.942(04) & \\
                           & 1.20          & 0.894(06) & 0.894(06) & \\
\hline
                           & 1.00          & 1.000(00) & 1.000(00) & \\
                           & 1.03          & 0.980(02) & 0.980(02) & \\
$B\rightarrow B$           & 1.05          & 0.969(03) & 0.969(03) & \\
                           & 1.10          & 0.940(05) & 0.940(05) & \\
                           & 1.20          & 0.882(09) & 0.882(09) & \\
\hline
                           & 1.00          & 1.025(17) & 1.013(03) & -0.020(37) \\
                           & 1.03          & 1.009(14) & 0.992(03) & -0.032(29) \\
$B\rightarrow D$           & 1.05          & 1.000(13) & 0.980(04) & -0.040(23) \\
                           & 1.10          & 0.976(11) & 0.953(04) & -0.048(17) \\
                           & 1.20          & 0.929(09) & 0.903(06) & -0.053(12) \\
\end{tabular}
\end{ruledtabular}
\caption{\label{tab:s0results}
Small volume results, $L_0=0.4$~fm. Results corresponding to the definition $D1$.}
\end{table}
\begin{table}
\begin{ruledtabular}
\begin{tabular}{lccccc}
 & $\beta$ & $T\times L^3$ & $N_{cnfg}$ & $k$ & $\theta$\\
\hline
$L_0a$ & 6.737 & $24\times 12^3$   & 608  & 0.12490 & 0.000000\\
       &       &                   &      & 0.12600 & 0.953456\\
       &       &                   &      & 0.12770 & 1.201080\\
       &       &                   &      & 0.12979 & 1.719172\\
       &       &                   &      & 0.13015 & 2.488491\\
       &       &                   &      & 0.13430 &\\
       &       &                   &      & 0.13460 &\\
       &       &                   &      & 0.13490 &\\
\hline
$L_0b$ & 6.420 & $16\times  8^3$   & 800  & 0.120674 & 0.000000\\
       &       &                   &      & 0.122220 & 0.953456\\
       &       &                   &      & 0.124410 & 1.201079\\
       &       &                   &      & 0.127985 & 1.719172\\
       &       &                   &      & 0.128066 & 2.488491\\
       &       &                   &      & 0.134304 &\\
       &       &                   &      & 0.134770 &\\
       &       &                   &      & 0.135221 &\\
\hline
$L_1A$ & 6.737 & $48\times 24^3$   & 260  &  0.12490 & 0.000000\\
       &       &                   &      &  0.12600 & 2.042983\\
       &       &                   &      &  0.12770 & 2.573569\\
       &       &                   &      &  0.12979 & 3.438340\\
       &       &                   &      &  0.13015 & 4.976980\\
       &       &                   &      &  0.13430 &\\
       &       &                   &      &  0.13460 &\\
       &       &                   &      &  0.13490 &\\
\hline
$L_1B$ & 6.420 & $32\times 16^3$   & 350  &  0.120674 & 0.000000\\
       &       &                   &      &  0.122220 & 2.042983 \\
       &       &                   &      &  0.124410 & 2.573569 \\
       &       &                   &      &  0.127985 & 3.438340 \\
       &       &                   &      &  0.128066 & 4.976980 \\
       &       &                   &      &  0.134304 & \\
       &       &                   &      &  0.134770 & \\
       &       &                   &      &  0.135221 & \\
\end{tabular}
\end{ruledtabular}
\caption{\label{tab:sims1}
Table of lattice simulations of the first step.}
\end{table}
\begin{table}
\begin{ruledtabular}
\begin{tabular}{ccccc}
$i\rightarrow f$ & $w$ & $\sigma_G$ & $\sigma_+$ & $\sigma_-$ \\ 
\hline
                           & 1.00          & 1.000(00) & 1.000(00) & \\
                           & 1.03          & 0.999(01) & 0.999(01) & \\
$D\rightarrow D$           & 1.05          & 0.997(02) & 0.997(02) & \\
                           & 1.10          & 0.993(03) & 0.993(03) & \\
                           & 1.20          & 0.985(05) & 0.985(05) & \\
\hline
                           & 1.00          & 1.000(00) & 1.000(00) & \\
                           & 1.03          & 0.997(02) & 0.997(02) & \\
$B\rightarrow B$           & 1.05          & 0.996(02) & 0.996(02) & \\
                           & 1.10          & 0.991(04) & 0.991(04) & \\
                           & 1.20          & 0.981(09) & 0.981(09) & \\
\hline
                           & 1.00          & 1.002(02) & 1.003(01) & 0.89(16)\\
                           & 1.03          & 0.999(03) & 1.000(02) & 0.86(13)\\
$B\rightarrow D$           & 1.05          & 0.996(04) & 0.998(02) & 0.83(11)\\
                           & 1.10          & 0.991(04) & 0.993(03) & 0.77(08)\\
                           & 1.20          & 0.980(05) & 0.983(05) & 0.72(08)\\
\end{tabular}
\end{ruledtabular}
\caption{\label{tab:s1results}
First step, from $L_0=0.4$~fm to $L_1=0.8$~fm. Results corresponding to the definition $D1$.}
\end{table}
\begin{table}
\begin{ruledtabular}
\begin{tabular}{lccccc}
 & $\beta$ & $T\times L^3$ & $N_{cnfg}$ & $k$ & $\theta$\\
\hline
$L_1a$ & 6.420 & $32\times 16^3$   & 360  & 0.126600 & 0.000000\\
       &       &                  &       & 0.127400 & 1.603930\\
       &       &                  &       & 0.128030 & 2.080840\\
       &       &                  &       & 0.128650 & 2.978423\\
       &       &                  &       & 0.129500 & 4.311249\\
       &       &                  &       & 0.134304 & \\
       &       &                  &       & 0.134770 & \\
       &       &                  &       & 0.135221 & \\
\hline
$L_1b$ & 5.960 & $16\times  8^3$   & 480  & 0.118128 & 0.000000\\
       &       &                   &      & 0.119112 & 1.603930 \\
       &       &                   &      & 0.120112 & 2.080840 \\
       &       &                   &      & 0.121012 & 2.978423 \\
       &       &                   &      & 0.122513 & 4.311249 \\
       &       &                   &      & 0.131457 & \\
       &       &                   &      & 0.132335 & \\
       &       &                   &      & 0.133226 & \\
\hline
$L_2A$ & 6.420 & $48\times 24^3$   & 250  &  0.126600 & 0.000000\\
       &       &                   &      &  0.127400 & 2.405895\\
       &       &                   &      &  0.128030 & 3.121260\\
       &       &                   &      &  0.128650 & 4.467634\\
       &       &                   &      &  0.129500 & 6.200000\\
       &       &                   &      &  0.134304 & \\
       &       &                   &      &  0.134770 & \\
       &       &                   &      &  0.135221 & \\
\hline
$L_2B$ & 5.960 & $24\times 12^3$   & 592  &  0.118128 & 0.000000\\
       &       &                   &      &  0.119112 & 2.405895\\
       &       &                   &      &  0.120112 & 3.121260\\
       &       &                   &      &  0.121012 & 4.467634\\
       &       &                   &      &  0.122513 & 6.200000\\
       &       &                   &      &  0.131457 & \\
       &       &                   &      &  0.132335 & \\
       &       &                   &      &  0.133226 & \\
\end{tabular}
\end{ruledtabular}
\caption{\label{tab:sims2}
Table of lattice simulations of the second step.}
\end{table}
\begin{table}
\begin{ruledtabular}
\begin{tabular}{ccccc}
$i\rightarrow f$ & $w$ & $\sigma_G$ & $\sigma_+$ & $\sigma_-$ \\ 
\hline
                           & 1.00          & 1.000(00) & 1.000(00) & \\
                           & 1.03          & 0.993(01) & 0.993(01) & \\
$D\rightarrow D$           & 1.05          & 0.988(02) & 0.988(02) & \\
                           & 1.10          & 0.973(05) & 0.973(05) & \\
                           & 1.20          & 0.921(16) & 0.921(16) & \\
\hline
                           & 1.00          & 1.000(00) & 1.000(00) & \\
                           & 1.03          & 0.992(02) & 0.992(02) & \\
$B\rightarrow B$           & 1.05          & 0.986(03) & 0.986(03) & \\
                           & 1.10          & 0.961(10) & 0.961(10) & \\
                           & 1.20          & 0.890(24) & 0.890(24) & \\
\hline
                           & 1.00          & 1.000(01) & 1.000(01) & 0.63(32)\\
                           & 1.03          & 0.992(02) & 0.993(02) & 0.65(25)\\
$B\rightarrow D$           & 1.05          & 0.987(02) & 0.988(02) & 0.67(21)\\
                           & 1.10          & 0.972(05) & 0.972(05) & 0.65(21)\\
                           & 1.20          & 0.921(14) & 0.921(14) & 0.45(33)\\
\end{tabular}
\end{ruledtabular}
\caption{\label{tab:s2results}
Second step, from $L_1=0.8$~fm to $L_2=1.2$~fm. Results corresponding to the definition $D1$.}
\end{table}
%

\end{document}